\documentclass[twocolumn,twoside,slac_two]{revtex4}
\usepackage{graphicx}
\usepackage{fancyhdr}
\usepackage{hyperref}
\hypersetup{
    colorlinks=true,
    urlcolor=magenta,
}
\begin{document}

\title{Influence of wildfires in Yakutia on interannual variability of AOT on measurements at stations near Yakutsk}

%

\author{S. Knurenko, I. Petrov}
\affiliation{Yu. G. Shafer Institute of Cosmophysical Research and Aeronomy SB RAS, Yakutsk, Russia}

\begin{abstract}
This paper presents data on the optical thickness AOT in the context of long-term observations, including in times of large-scale forest fires in Yakutia. Sudden changes in such features as the AOT point directly to the impact of fires on aerosol structure of the atmosphere during the summer.

\end{abstract}

\maketitle

\thispagestyle{fancy}


\section{Introduction}\label{PetrovIS-intro}

Polar regions of Yakutia, which can be attributed to the neighborhoods of Yakutsk (N 61$^\circ$39$^\prime$, E 129$^\circ$22$^\prime$, Alt 118 m), from the geophysics point of view, is interesting with anomalous manifestations in the upper atmosphere above 80 km above sea level: the invasion of flux of electrons ( auroras, radio noises etc.), their interactions with the environment, with the loss of significant energy and further the mechanism or mechanisms of energy transfer in the lower atmosphere. To some extent, these processes together with galactic cosmic rays may be involved in the formation of weather in the world ~\cite{PetrovIS-bibref1, PetrovIS-bibref2}. In ~\cite{PetrovIS-bibref3, PetrovIS-bibref4}, it was considered that the effect of solar and galactic cosmic rays on the temperature, therefore the weather and climate of our planet is practically proven.

\begin{figure}
\includegraphics[width=0.8\linewidth]{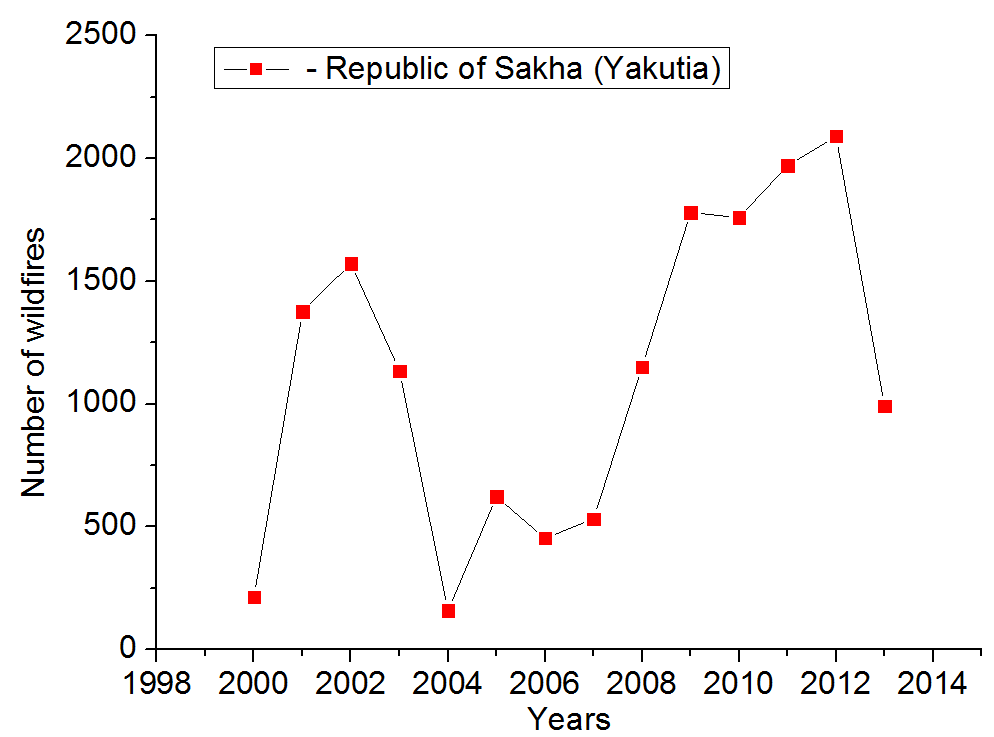}
\caption{Number of wildfires in Yakutsk by years}
\label{PetrovIS-fig1}
\end{figure}

Observation of the atmosphere of the Yakutsk array are carried out continuously for more than 45 years. During this time a large amount of data collected by such characteristics as temperature, pressure and humidity of the surface air and troposphere section ~\cite{PetrovIS-bibref5}. Data and spectral transparency of the atmosphere is also collected at a wavelength $\lambda$ = 430 nm, which are derived from the relative frequency of the EAS with energies 10$^{15}$ - 10$^{16}$ eV ~\cite{PetrovIS-bibref6}. All this made a database of meteorological parameters of the atmosphere, which is used for analysis of air showers, as well as analysis of time series to study the physical characteristics of the atmosphere and the communication of its response to the geophysical processes occurring in the polar zone of Yakutia ~\cite{PetrovIS-bibref5}. For example, a temperature trend for the winter in Yakutsk was found that equals to (0.6 $\pm$ 0.2) $^\circ$C / 10 years. Analysis of the data also indicates that the atmosphere is influenced by such factors as the man-made disasters, volcanic activity in Kamchatka, global wildfires and the greenhouse effect. Warming pointed to our data, manifested everywhere in Yakutia: in winter at times significant reduction in days with anomalously low temperatures in the summer dry and sultry weather, resulting in a significant increase in forest fires (Fig.~\ref{PetrovIS-fig1}).

\begin{figure}
\includegraphics[width=0.8\linewidth]{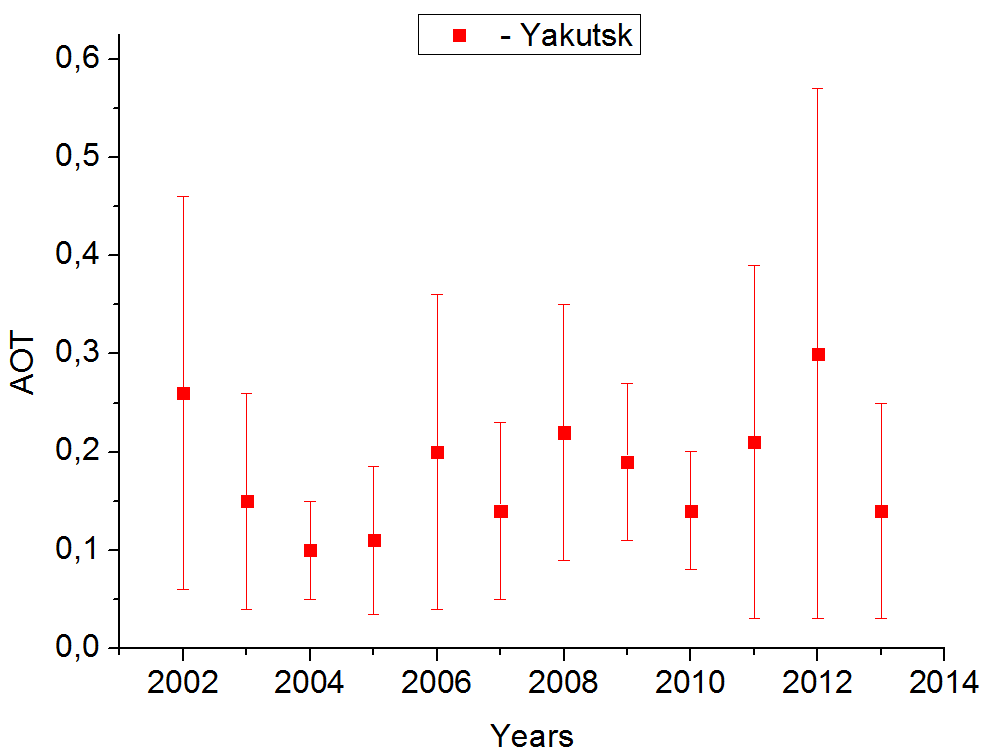}
\caption{The interannual variability of AOT in the Yakutsk region. We consider the observation period from 2002 to 2013}
\label{PetrovIS-fig2}
\end{figure}

\section{Atmospheric Optical Thickness}

Aerosol atmosphere - a suspension of solid and liquid microparticles which are classified as the height and on the geographic area ~\cite{PetrovIS-bibref7}. This classification allows you to select specific sources of atmospheric aerosol and aerosol basic transformation processes under the influence of geophysical factors, including cosmic radiation and galactic origin ~\cite{PetrovIS-bibref3, PetrovIS-bibref2}.

Yakutia is located in the area with an extreme continental climate, the temperature in winter -55 $^\circ$C and in summer 35 $^\circ$C ~\cite{PetrovIS-bibref8}. As a result, during the winter months there is a temperature inversion, which leads to an increase in air density in the surface layers of the atmosphere, that leads to the frosty mists ~\cite{PetrovIS-bibref5}. In the summer months due to the high temperatures and low rainfall, frequent forest fires, caused by dry thunderstorms, due to which increases the AOT. Figure ~\ref{PetrovIS-fig2} shows the data for the parameter AOT, averaged monthly during the observation period from 2002 to 2013. The figure shows that there were anomalous emissions in 2002 and 2012.

\section{Abnormal 2002 and 2012 years}

\begin{figure}
\includegraphics[width=0.8\linewidth]{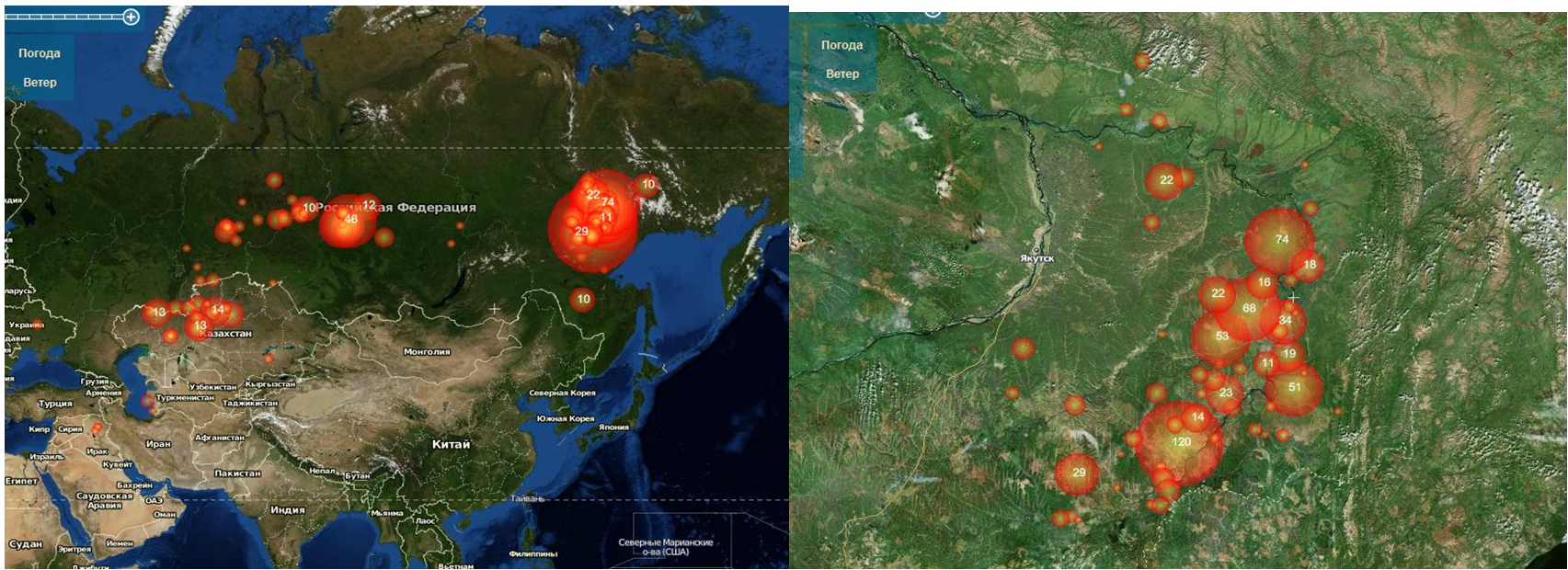}
\caption{The number of recognized wildfires in Western Siberia and Yakutia in the summer months of 2012. Showing reconstructed satellite image}
\label{PetrovIS-fig3}
\end{figure}

In spring and summer, there are wildfires in Yakutia with different intensity, which largely affect both the average monthly and annual value of AOT. This is - first of all, refers to the South and Central Yakutia, including the city of Yakutsk. As can be seen from the data shown in Fig. ~\ref{PetrovIS-fig2}, interannual value of AOT varies from 0.1 in calm years and up to 0.3 in the years with an abnormally large number of wildfires. According to ~\cite{PetrovIS-bibref9} and the data given in Fig. ~\ref{PetrovIS-fig2}, large-scale wildland fire were reported in 2002 and 2012, which affected the annual AOT, the average value, which reached a value equal to 0.26$\pm$0.20 in 2002 and 0.29$\pm$0.25 in 2012. Here, of course, stands out in 2012, when the value of AOT is three times higher than the average of other years. Fig. ~\ref{PetrovIS-fig3} shows the area with the largest number of fires, from space. It provided two regions: Western Siberia, Tomsk and Novosibirsk region and Eastern Siberia, South and Central Yakutia. The most intense fires marked in Yakutia. In practice, in the absence of wind direction, I could settled for a long time in these areas, and significantly worsened the aerosol composition of the air.

\begin{figure}
\includegraphics[width=0.8\linewidth]{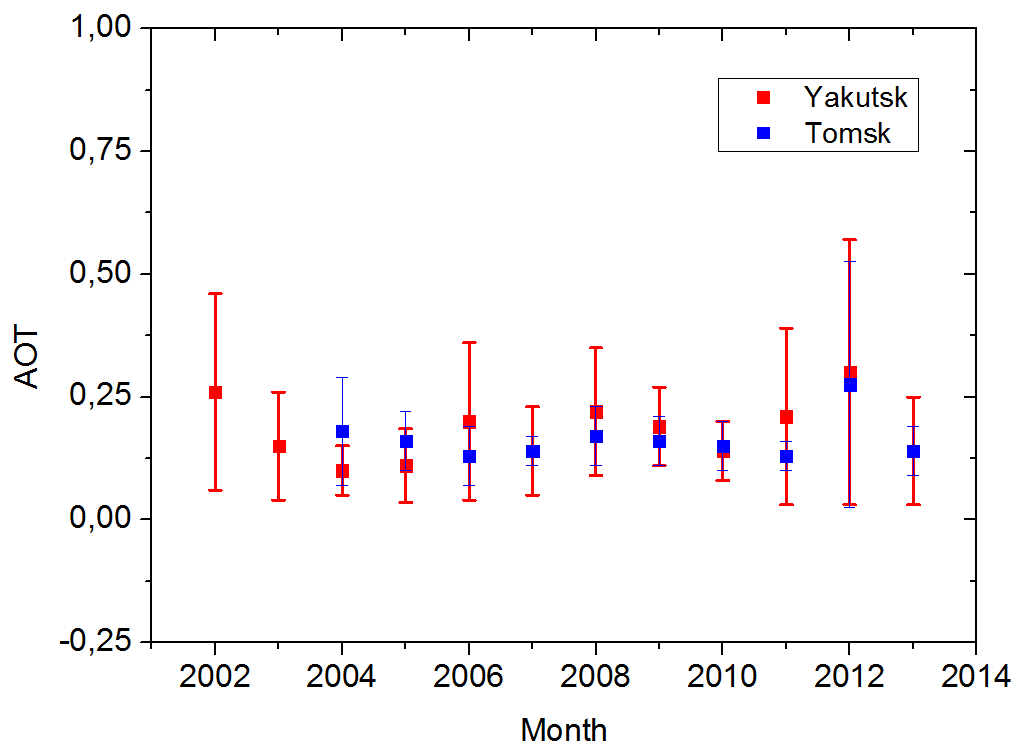}
\caption{Comparison of long-term data on the characteristics from the measured in Yakutsk and Tomsk in 2002 to 2013 time period}
\label{PetrovIS-fig4}
\end{figure}

Figure ~\ref{PetrovIS-fig4} shows a comparison of the data obtained from measurements of AOT at the city of Tomsk region ~\cite{PetrovIS-bibref10} and at the city of Yakutsk ~\cite{PetrovIS-bibref11}. A large number of wildfires in South - East of Yakutia in the summer months (Fig. ~\ref{PetrovIS-fig3}) has led to a large-scale air pollution. A similar pattern was observed in 2012, in Western Siberia (Fig. ~\ref{PetrovIS-fig3}), in the areas of the city of Tomsk and Novosibirsk. According to the data shown in Fig. ~\ref{PetrovIS-fig4}, the value of AOT, independently measured in Tomsk and Yakutsk coincide with each other. The observed excess of the average for the year of AOT in 2002 and 2012 by measurements near Yakutsk in 2012, Tomsk region, due to the intense fires in these years. This correlation indicates a strong influence of anthropogenic factors on the surface atmosphere.

\section{Conclusion}

Global warming can be traced throughout Yakutia, including the polar zone. This leads to a numerous wildfires, especially during the summer, which affects the "purity" of the atmosphere. For example, in 2012 due to large-scale wildland fires in Western and Eastern Siberia anomalies in the AOT values from measurements in Tomsk and Yakutsk were observed. As you can see, one of the essential factors of interannual variability of AOT in this period are large-scale wildfires, which resulted in the atmosphere comes a large number of fine smoke aerosols. Fig. 4 shows that in the Yakutsk region there was an almost threefold increase in AOT ($\sim$0,3) with a multi-year norm ($\sim$0,13). Thus, the main role in the interannual variations of AOT in Central Yakutia, except for seasonal circulation of air masses, plays a fine spray from wildfires, which largely affects the average annual value of AOT.
Therefore, we can conclude that the long-term effects of fires on the atmosphere can lead to disastrous consequences for the population of the regions, and cause irreparable damage to the surrounding nature.

\bigskip 

\begin{thebibliography}{0}
\expandafter\ifx\csname natexlab\endcsname\relax\def\natexlab#1{#1}\fi
\expandafter\ifx\csname bibnamefont\endcsname\relax
  \def\bibnamefont#1{#1}\fi
\expandafter\ifx\csname bibfnamefont\endcsname\relax
  \def\bibfnamefont#1{#1}\fi
\expandafter\ifx\csname citenamefont\endcsname\relax
  \def\citenamefont#1{#1}\fi
\expandafter\ifx\csname url\endcsname\relax
  \def\url#1{\texttt{#1}}\fi
\expandafter\ifx\csname urlprefix\endcsname\relax\def\urlprefix{URL }\fi
\providecommand{\bibinfo}[2]{#2}
\providecommand{\eprint}[2][]{\url{#2}}

\end{thebibliography}


\begin{thebibliography}{9}   

\bibitem{PetrovIS-bibref1}
G.A. Jerebsov, V.A. Kovalenko, S.I.Molodych. 'Optics of the Atmosphere and Ocean', \textbf{21}, N 1, 53-59, 2008.
\bibitem{PetrovIS-bibref2}
G.F. Krymsky. 'Solar Terrestrial Physics', \textbf{9}, 44-46, 2006.
\bibitem{PetrovIS-bibref3}
V.A. Dergachev. 'Conference Proc.' \textbf{2}, 452-455, 2003.
\bibitem{PetrovIS-bibref4}
M.I. Pudovkin, O.M. Raspopov. 'Geomagnetism and Aeronomy'. \textbf{32}, N 5, 1-22, 1992.
\bibitem{PetrovIS-bibref5}
S.P. Knurenko, I.S. Petrov. 'Proc. SPIE 9292, 20th International Symposium on Atmospheric and Ocean Optics: Atmospheric Physics', 92925B, 2014.
\bibitem{PetrovIS-bibref6}
M.N. Dyakonov, S.P. Knurenko, V.A. Kolosov, I.Ye. Sleptsov. 'Optics of atmosphere', \textbf{4}, 868-873, 1991.
\bibitem{PetrovIS-bibref7}
Yu.E. Heinz, A.A. Zemlyanov, V.Ye. Zuev et al. 'Non-linear aerosol optics of the atmosphere', p.256, 1999.
\bibitem{PetrovIS-bibref8}
S.M. Sakerin, D.M. Kabanov, M.V. Panchenko et al. 'Optics of the Atmosphere and Ocean' \textbf{18}, N 11, 968-975, 2005.
\bibitem{PetrovIS-bibref9}
O.A. Tomshin, V.S. Solovyev. 'XIX International Symposium on Atmospheric and Ocean', p. 110, 2013.
\bibitem{PetrovIS-bibref10}
S.M. Sakerin, D.M. Kabanov. 'XIX International Symposium on Atmospheric and Ocean', 324-327, 2013.
\bibitem{PetrovIS-bibref11}
http: // aeronet.gshc.nasa.gov./
\end{thebibliography}

\end{document}